\documentclass[final,5p,times,twocolumn]{elsarticle}

\usepackage{graphicx}

\usepackage{bbm, bm, mathrsfs, amsthm, amsmath, amssymb}

\usepackage{hyperref}

\journal{Physics Letters A}

\begin{document}

\begin{frontmatter}



\title{Quantum cube: A toy model of a qubit}


\author{Pawel Blasiak}

\address{Institute of Nuclear Physics, Polish Academy of Sciences\\
ul.~Radzikowskiego~152, PL 31-342 Krak\'ow, Poland}
\ead{Pawel.Blasiak@ifj.edu.pl}
\ead[url]{www.ifj.edu.pl/~blasiak}

\begin{abstract}
	Account of a system may depend on available methods of gaining information.
	We discuss a simple discrete system whose description is affected by a specific model of measurement and transformations. It is shown that the limited means of investigating the system make the epistemic account of the model indistinguishable from a constrained version of a qubit corresponding to the convex hull of eigenstates of Pauli operators, Clifford transformations and Pauli observables.
\end{abstract}

\begin{keyword}
Epistemic view of quantum theory\sep Toy model of a qubit\sep Discrete systems



\end{keyword}

\end{frontmatter}





\section{Introduction}
Interest in toy models of quantum phenomena stems from simplicity of classical concepts they are built on. Besides intuitive insights, they also provide a convenient tool for studying the distinctive features of quantum theory.
Recent research show that many phenomena typically associated with strictly quantum mechanical effects have their analogues in the classical realm~\cite{Sp07,Ha99,Ki03,DaPlPl02,BaRuSp12}. These results suggest that quantum states can be seen as states of knowledge and considerable efforts are presently made towards understanding of possible $\psi$\emph{-epistemic} reconstructions of the theory~\cite{HaSp10,Ha04,Mo08,DaSuPaBr08,CoEdSp11,PuBaRu12}. Most notable in this respect is the Spekkens' toy model~\cite{Sp07} which reproduces a surprisingly wide range of quantum phenomena in a simple discrete system constrained by the so-called 'knowledge balance principle'. However, despite qualitative resemblance, it cannot be taken as a restricted version of the theory. In particular, on top of being local and non-contextual, the model fails to properly account for transformation properties of a single qubit. Sources of these dissimilarities have been recently well explored~\cite{SkRoSa08,CoEd11,Pu12,WaBa12}, as well as some other examples have been proposed~\cite{BaRuSp12,La12,En07}.

In this Letter, we focus on the case of a single qubit for which a simple discrete system is given faithfully reproducing the characteristic quantum-like behavior of a certain nontrivial subset of states of a qubit. 
In the model we explicitly describe what constitutes a measurement and how the system can be transformed, thereby restricting an agent's capabilities of gaining information about the system. Careful analysis of the model will show that agent's account of the system is fully equivalent to a constrained version of a qubit corresponding to the convex hull of eigenstates of Pauli operators, Clifford transformations and Pauli observables.

\section{Conventional probability setup}

\begin{figure}[h!]
\begin{center}
\includegraphics[scale=0.16]{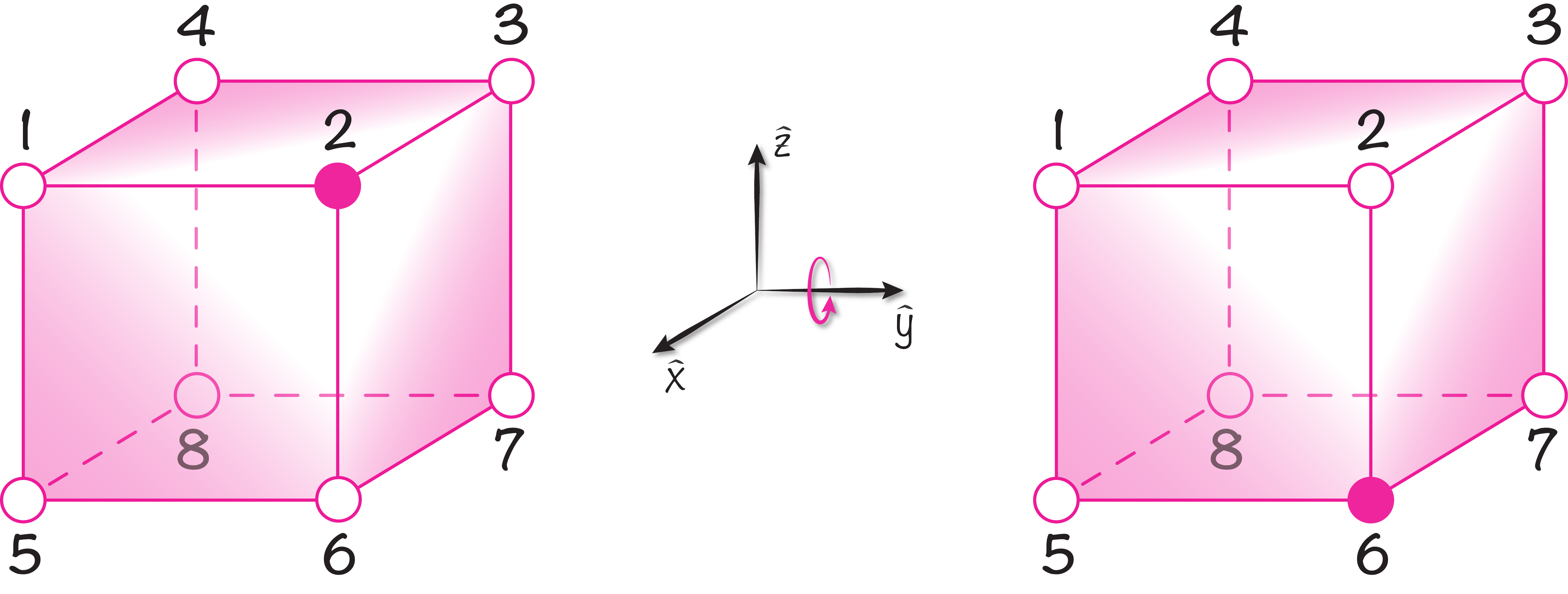}
\end{center}\vspace{-0.3cm}
\caption{\label{Fig1} Eight vertices of a cube represent ontic states of a system. The system in state $\omega_2$ (on the left) can be transformed to state $\omega_6$ (on the right) e.g. via rotation $R_{\hat{y}}(\tfrac{\pi}{2}$).}
\end{figure}

Let us consider an elementary system with $8$ possible (ontic) states $\Omega=\{\omega_1,...,\omega_8\}$. For future convenience, dictated by a natural link to geometry and symmetry under $\pi/2$ rotations, we will represent the state space $\Omega$ as a cube and depict the system in state $\omega_i$ as occupying the $i$-th vertex, see Fig.\ref{Fig1}.
A standard probabilistic description consists in specifying the probability vector $\bm{p}=(p_1,...,p_8)^T$. It belongs to a convex set $\Delta=\{\,(p_1,...,p_8)^T:\sum_{i=1}^8\,p_i=1\,,\,p_i\geqslant0\,\}$ spanned by extremal states $\bm{p}_1,...,\bm{p}_8$ (where $(\bm{p}_i)_j=\delta_{ij}$) corresponding to the system definitely being in the ontic state $\omega_1,...,\omega_n$ respectively.
Vector $\bm{p}$ encapsulates all information about the system and probability $p_i$ of finding it in state $\omega_i$ is readily given by $P_{\bm{p}}(i)=\bm{p}_i\cdot\bm{p}$.

From the 'ontic' standpoint any vector $\bm{p}\in\Delta$ is a valid probability state. In particular, one requires that the system can be prepared in a definite ontic state $\omega_i$.
However, in reality such a privileged situation may not obtain, with an agent facing limitations on measuring, preparing and transforming the system. Below, we discuss a simple model in which constrained measurement and transformation procedures modify the conventional picture compelling agent to adopt essentially 'epistemic' -- rather than 'ontic' -- description of the system.

\begin{figure*}[t!]
\begin{center}
\includegraphics[scale=0.115]{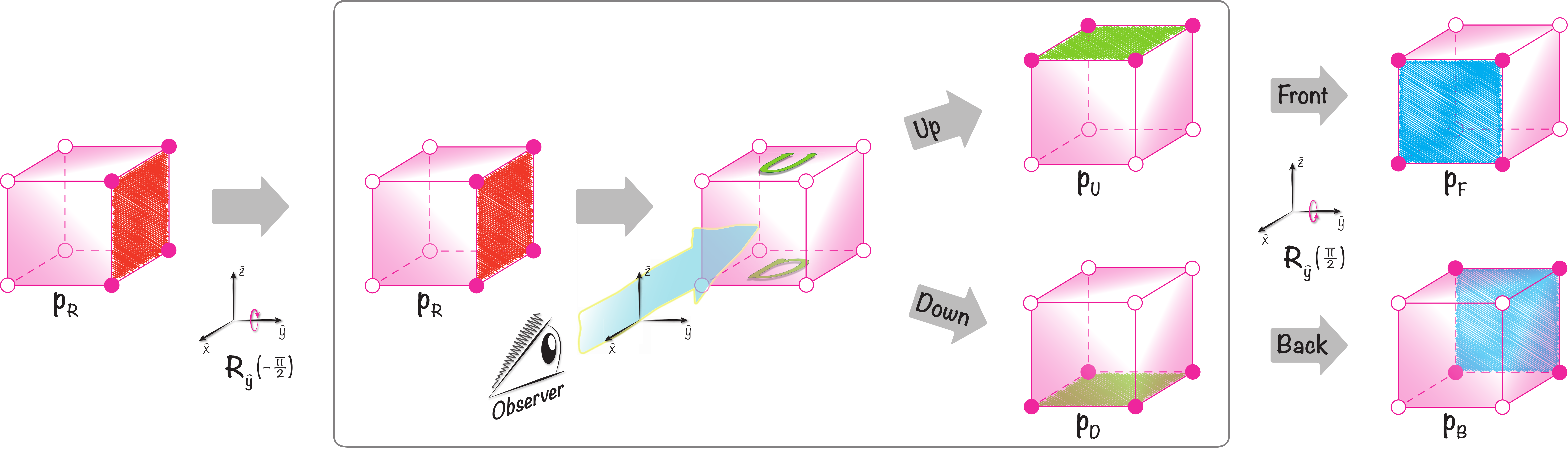}
\end{center}\vspace{-0.4cm}
\caption{\label{Fig2}Box in the middle illustrates the basic measurement $M_\mathbbm{1}$ which distinguishes only between the upper $U$ and the lower $L$ face of the cube and produces states $\bm{p}_{\scriptscriptstyle{U}}$ and $\bm{p}_{\scriptscriptstyle{D}}$ respectively. The whole picture demonstrates effect of the measurement procedure $M_T=T\circ M_{\mathbbm{1}}\circ T^{-1}$ for $T=R_{\hat{y}}(\tfrac{\pi}{2})$ which discriminates between the front $F$ and the back $B$ face of the cube. Shaded faces of the cube depict equiprobable mixtures of the adjacent vertices representing ontic states of the system.}\vspace{-0.3cm}
\end{figure*}
\section{Epistemic Model}

\subsection{Model of Measurement}

A primary role of measurement is to reveal information. Let a \emph{basic measurement}, denoted by $M_\mathbbm{1}$, be defined to answer the question whether a given elementary system is in one of the upper $U=\{\omega_1,...,\omega_4\}$ or lower $D=\{\omega_5,...,\omega_8\}$ states, and subsequently alter the system leaving it with equal probabilities in one of the four compatible states, i.e.
\begin{eqnarray}\nonumber
\omega_1,...,\omega_4&\stackrel{\text{\textsc{Up}}}{\longrightarrow}&\bm{p}_{\scriptscriptstyle{U}}=\left(\tfrac{1}{4},\tfrac{1}{4},\tfrac{1}{4},\tfrac{1}{4},0,0,0,0\right)^T\,,\\\nonumber
\omega_5,...,\omega_8&\stackrel{\text{\textsc{Down}}}{\longrightarrow}&\bm{p}_{\scriptscriptstyle{D}}=\left(0,0,0,0,\tfrac{1}{4},\tfrac{1}{4},\tfrac{1}{4},\tfrac{1}{4}\right)^T\,.
\end{eqnarray}
See Fig.\ref{Fig2} (in the box) for a schematic illustration. Observe that in spite of disturbance this definition guarantees reproducibility of results on individual systems.
Informally, the measurement can be imagined as if an observer was looking at the cube representing the system from the $\hat{x}$ direction distinguishing only between states on the upper and lower face, and at the same time jiggling the cube in the horizontal plane. In short, the measurement discriminates states in the $\hat{z}$ direction while randomizing the system in the $\hat{x}$-$\hat{y}$ plane.

Clearly, information gain in such a measurement comes at a cost of disturbance.
Measurement on a system randomly chosen from an ensemble described by $\bm{p}$ will yield outcome $\mu=U,D$ with probability given by the formula $P_{\bm{p}}(\mu)=4\ \bm{p}_\mu\cdot\bm{p}$. In consequence of the measurement (performed on each element of the the ensemble) the state changes to the mixture $\bm{p}\rightarrow P_{\bm{p}}(U)\,\bm{p}_{\scriptscriptstyle{U}}+P_{\bm{p}}(D)\,\bm{p}_{\scriptscriptstyle{D}}$. Effectively it is a stochastic transformation which in a compact notation reads $\bm{p}\rightarrow M_{\mathbbm{1}}\,\bm{p}$, where $M_{\mathbbm{1}}$ is a block diagonal matrix with two blocks of size four entirely filled with $\tfrac{1}{4}$'s, i.e. $(M_{\mathbbm{1}})_{ij}=\tfrac{1}{4}\,[i,j\leqslant4]+\tfrac{1}{4}\,[i,j>4]$ in Iverson notation.
 

Now suppose that the system can be probed \textit{only} with the measurement $M_\mathbbm{1}$ and \textit{no} additional transformations are available. Then the only states within agent's reach for preparing the system are $\bm{p}_{\scriptscriptstyle{U}}$ and $\bm{p}_{\scriptscriptstyle{D}}$, thereby rendering $\Delta_0=\{\,\alpha\,\bm{p}_{\scriptscriptstyle{U}}+\beta\,\bm{p}_{\scriptscriptstyle{D}}:\alpha+\beta=1,\,\alpha,\beta\geqslant0\,\}$ to be a maximal set of states at her disposal. That being so, the state space in this scenario is equivalent to a \emph{classical bit} with $\bm{p}_{\scriptscriptstyle{U}}$ and $\bm{p}_{\scriptscriptstyle{D}}$ taking the role of binary states 0 and 1
(note that these states remain unaltered by the measurement). We stress the fact that from the agent's perspective it is a complete description of the system perceived and tackled with her limited resources. This is to say that information coded by vectors $\bm{p}\in\Delta_0$ 
is just enough to account for all possible actions in hand, i.e. preparations (mixtures of measurement outputs), transformations (only the identity transformation $\mathbbm{1}$) and measurements (only measurement $M_{\mathbbm{1}}$). 
Moreover, as long as these cognitive restrictions apply, the agent has not the least hint of a more complex nature of the system which appears to her indistinguishable from a classical bit. It will be shown to expose more structure of the ontic state space and reshape the epistemic picture to a form analogous to a certain set of states of a qubit.

\subsection{Structure of transformations}

Let us extend the above setup to include rotations of the cube through angle $\tfrac{\pi}{2}$ about axes $\hat{x}$, $\hat{y}$, $\hat{z}$ and combinations thereof (see Fig.\ref{Fig1}). Here, by rotation of the system we mean the associated permutation of its ontic state space $\Omega$, e.g. $R_{\hat{x}}(\tfrac{\pi}{2})$, $R_{\hat{y}}(\tfrac{\pi}{2})$, $R_{\hat{z}}(\tfrac{\pi}{2})$ represent permutations $(1562)(3487)$, $(1584)(2673)$, $(1234)(5678)$ respectively. The full set of such \emph{transformations}, denoted further by $\mathcal{R}_{\pi/2}$, consists of 24 elements which form a group of rotational symmetries of a cube~\cite{Ar88}.
This explains our initial choice of a cube for a geometrical representation of the state space of the system -- its symmetries aptly capture properties of the transformation set $\mathcal{R}_{\pi/2}$. Let us note in passing that $\mathcal{R}_{\pi/2}$ is isomorphic to the group of Clifford transformations of a qubit.
Of course, $\mathcal{R}_{\pi/2}$ is still a limited set of transformations if compared with all conceivable mappings, yet it is more than in the foregoing discussion of the basic measurement $M_{\mathbbm{1}}$ with the trivial set of transformations $\mathcal{R}_0=\{\mathbbm{1}\}$. 
Extension of the allowable set of transformations (from $\mathcal{R}_0$ to $\mathcal{R}_{\pi/2}$) gives an edge in probing the system which comes from the fact that measurements and preparations may, now, combine with transformations. Below we describe the improved toolkit at agents disposal and discuss the emerging epistemic picture of the system.

\begin{figure}[t]
\begin{center}
\includegraphics[scale=0.2]{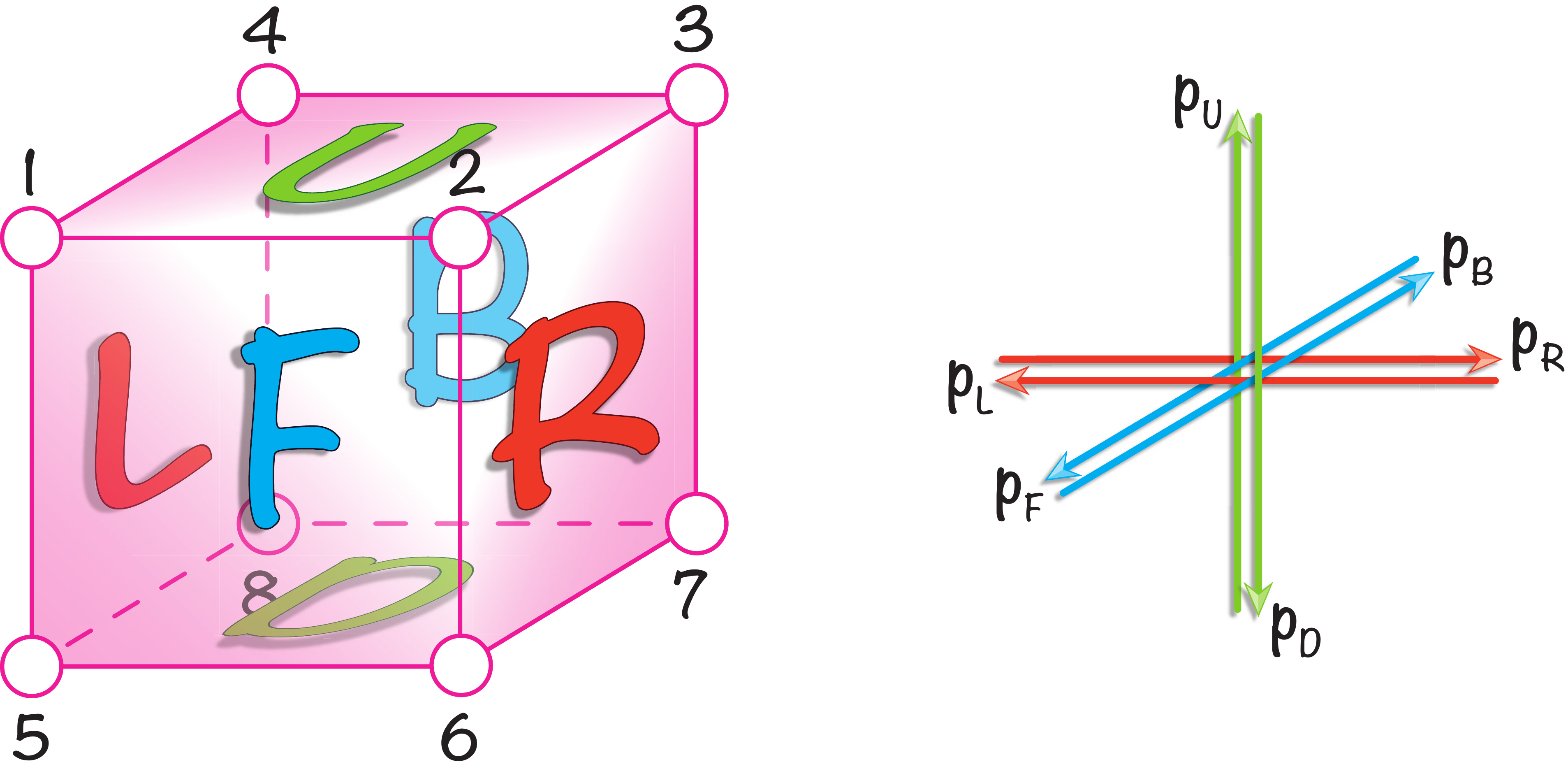}
\end{center}\vspace{-0.3cm}
\caption{\label{Fig3} On the left, faces of the cube corresponding to states $\bm{p}_{\scriptscriptstyle{U}}$, $\bm{p}_{\scriptscriptstyle{D}}$, $\bm{p}_{\scriptscriptstyle{L}}$, $\bm{p}_{\scriptscriptstyle{R}}$, $\bm{p}_{\scriptscriptstyle{F}}$, $\bm{p}_{\scriptscriptstyle{B}}$. On the right, vectors representing six pairs of opposite faces tested by general measurement procedures.}\vspace{-0.4cm}
\end{figure}

Firstly, the system prepared as the outcome of measurement $M_{\mathbbm{1}}$ in state $\bm{p}_{\scriptscriptstyle{U}}$ or $\bm{p}_{\scriptscriptstyle{D}}$ can be transformed to one of the four new states:
\begin{eqnarray}\nonumber
\bm{p}_{\scriptscriptstyle{L}}=\left(\tfrac{1}{4},0,0,\tfrac{1}{4},\tfrac{1}{4},0,0,\tfrac{1}{4}\right)^T\!\!,
\ \ \bm{p}_{\scriptscriptstyle{R}}=\left(0,\tfrac{1}{4},\tfrac{1}{4},0,0,\tfrac{1}{4},\tfrac{1}{4},0\right)^T\!\!,
\\\nonumber
\bm{p}_{\scriptscriptstyle{F}}=\left(\tfrac{1}{4},\tfrac{1}{4},0,0,\tfrac{1}{4},\tfrac{1}{4},0,0\right)^T\!\!,
\ \ \bm{p}_{\scriptscriptstyle{B}}=\left(0,0,\tfrac{1}{4},\tfrac{1}{4},0,0,\tfrac{1}{4},\tfrac{1}{4}\right)^T\!\!.
\end{eqnarray}
These states correspond to the system being with equal probabilities in one of the ontic states  in the respective sets $L=\{\omega_1,\omega_4,\omega_5,\omega_8\}$, $R=\{\omega_2,\omega_3,\omega_6,\omega_7\}$, $F=\{\omega_1,\omega_2,\omega_5,\omega_6\}$, $B=\{\omega_3,\omega_4,\omega_7,\omega_8\}$. The sets $U$, $D$, $L$, $R$, $F$ and $B$ coincide with the faces of the cube (up, down, left, right, front and back), as depicted in Fig.\ref{Fig3} on the left.
Note that each state can be obtained in more than one way, e.g. $\bm{p}_{\scriptscriptstyle{F}}$ obtains from $\bm{p}_{\scriptscriptstyle{U}}$ by rotation $R_{\hat{y}}(\tfrac{\pi}{2})$ or  $R_{\hat{x}}(-\tfrac{\pi}{2})\circ R_{\hat{y}}(\tfrac{\pi}{2})$ (the latter is equivalent to the rotation through $\tfrac{2\pi}{3}$ about the diagonal joining vertices 1 and 7).
Since transformations from the set $\mathcal{R}_{\pi/2}$ just shuffle faces of the cube, then $\bm{p}_{\scriptscriptstyle{U}}$, $\bm{p}_{\scriptscriptstyle{D}}$, $\bm{p}_{\scriptscriptstyle{L}}$, $\bm{p}_{\scriptscriptstyle{R}}$, $\bm{p}_{\scriptscriptstyle{F}}$ and $\bm{p}_{\scriptscriptstyle{B}}$ is an exhaustive collection of states obtainable in this way. Hence, the most general states that can be prepared by the agent belong to the convex set
\begin{eqnarray}\label{StateSpace}
\Delta_{\pi/2}=\left\{\,{\sum}_{\mu}p_\mu\,\bm{p}_\mu:{\sum}_{\mu}\,p_\mu=1\,,\,p_\mu\geqslant0\right\}\,,
\end{eqnarray}
where $\mu=U,D,L,R,F,B$.

Secondly, each transformation $T\in \mathcal{R}_{\pi/2}$ has a canonical representation as 8x8 permutation matrix $T_{ij}=\delta_{i\,\sigma(j)}$, where $\sigma$ is the permutation of the vertices of the cube induced by the rotation $T$. This readily carries over into probabilistic description of the system whose states transform via the stochastic map
\begin{eqnarray}\label{Transformation}
\bm{p}\rightarrow T\,\bm{p}\,. 
\end{eqnarray}
Since $\Delta_{\pi/2}$ is mapped into itself by transformations in $\mathcal{R}_{\pi/2}$, we infer that $\Delta_{\pi/2}$ remains the maximal set of states available to the agent. 

Thirdly, richer set of transformations provides subtler means for probing the system. A measurement preceded by a transformation reveals different information than a bare measurement $M_{\mathbbm{1}}$. Therefore, for each $T\in \mathcal{R}_{\pi/2}$ we define a new measurement procedure
\begin{eqnarray}\label{Measurement}
M_T=T\circ M_{\mathbbm{1}}\circ T^{-1}\,,
\end{eqnarray}
which consists of the preparatory phase $T^{-1}$, the measurement $M_{\mathbbm{1}}$ furnishing the outcome, and the closing transformation preparing the output; see Fig.\ref{Fig2}.  
Although there are 24 such procedures (i.e. as many as there are transformations in $\mathcal{R}_{\pi/2}$), some of them are equivalent and we get only 6 essentially different measurements. This can be seen by observing that the upper $U$ and the lower $D$ face of the cube, being distinguished in measurement $M_{\mathbbm{1}}$, is now replaced by some other pair of opposite faces due to the preparatory rotation $T^{-1}$. Thus, measurement $M_{T}$ answers the question on which of these two faces the system resides; choice of the pair depending on $T$, e.g. $T=\mathbbm{1},\,R_{\hat{x}}(-\tfrac{\pi}{2}),\,R_{\hat{y}}(-\tfrac{\pi}{2}),\,R_{\hat{y}}(-\pi)$ test for $(U,D)$, $(L,R)$, $(F,B)$, $(D,U)$ respectively. Since $\mathcal{R}_{\pi/2}$ are rotational symmetries of the cube there are only 6 different measurements which distinguish between elements of the pairs $(U,D)$, $(D,U)$, $(L,R)$, $(R,L)$, $(F,B)$, $(B,F)$.
It is convenient to associate with these pairs vectors $\bm{p}_{\scriptscriptstyle{U}}$, $\bm{p}_{\scriptscriptstyle{D}}$, $\bm{p}_{\scriptscriptstyle{L}}$, $\bm{p}_{\scriptscriptstyle{R}}$, $\bm{p}_{\scriptscriptstyle{F}}$ and $\bm{p}_{\scriptscriptstyle{B}}$ respectively; see Fig.\ref{Fig3} on the right. Then, probability of outcome $\nu$ in the measurement $\bm{p}_\mu$ performed on a system described by $\bm{p}$ is given by the formula
\begin{eqnarray}\label{Probability}
P_{\bm{p}}(\nu)=4\ \bm{p}_\nu\cdot\bm{p}\,,
\end{eqnarray}
where $\nu$ takes only two values in accord with the chosen measurement $\bm{p}_\mu$, e.g. for $\mu=L$ which tests $(L,R)$ we have $\nu=L,R$. Note that the system gets altered in consequence of the measurement. Due to the closing part in the definition Eq.(\ref{Measurement}) it is left with equal probabilities in one of the four compatible states, i.e. the output state is $\bm{p}_\nu$ if the outcome was $\nu$; cf. Fig.\ref{Fig2}. 
We remark that in spite of disturbance measurements performed on individual systems are reproducible, i.e. subsequent measurement of the same kind yields the same result. Observe, however, that same measurements intervened by a measurement of another kind provide in general different results -- that being a sign of non-commutativity.



\subsection{Embedding in the Hilbert space}




To recap, we have seen how a constrained repertoire of available preparations, transformations and measurements set boundaries on a possible knowledge of the system. In our toy model the agent, whose basic toolkit consists only of measurement $M_{\mathbbm{1}}$ and transformations $\mathcal{R}_{\pi/2}$, is entirely confined to the subspace $\Delta_{\pi/2}$. This is to say that it provides just enough information which is needed to account for all her possible actions
and by no means can she reach beyond this set. Thus collection of states $\Delta_{\pi/2}$ together with specific rules of processing them provide a complete epistemic account of the system as seen from the agent's perspective.

Most interestingly, our toy model has a natural embedding in a familiar Hilbert space description of a qubit. To see this consider the following correspondence between the probability vectors $\bm{p}_\mu$ and the pure states of a qubit corresponding to eigenstates of Pauli operators:
\begin{eqnarray}\nonumber
\bm{p}_{\scriptscriptstyle{U}}\leftrightarrow |0\rangle\langle 0|
\ \ \ &
\bm{p}_{\scriptscriptstyle{F}}\leftrightarrow |+\rangle\langle +|
\ \ \ &
\bm{p}_{\scriptscriptstyle{R}}\leftrightarrow |+i\rangle\langle +i|
\\\nonumber
\bm{p}_{\scriptscriptstyle{D}}\leftrightarrow |1\rangle\langle 1|
\ \ \ &
\bm{p}_{\scriptscriptstyle{B}}\leftrightarrow |-\rangle\langle -|
\ \ \ &
\bm{p}_{\scriptscriptstyle{L}}\leftrightarrow |-i\rangle\langle -i|
\end{eqnarray}
where $|\pm\rangle=\tfrac{1}{\sqrt{2}}\left(|0\rangle\pm|1\rangle\right)$ and  $|\pm i\rangle=\tfrac{1}{\sqrt{2}}\left(|0\rangle\pm i|1\rangle\right)$. This furnishes a one-to-one mapping from $\Delta_{\pi/2}$ into the following subset of density operators
\begin{eqnarray}\label{StateSpaceHilbert}
\Xi=\left\{\,{\sum}_{\kappa}p_\kappa\,|\kappa\rangle\langle\kappa|:{\sum}_{\kappa}\,p_\kappa=1\,,\,p_\kappa\geqslant0\,\right\}\,,
\end{eqnarray}
where $\kappa=0,1,\pm,\pm i$. Explicitly, it is defined by
\begin{eqnarray}\nonumber
\Delta_{\pi/2}\ni{\sum}_\mu\,p_\mu\,\bm{p}_\mu=\bm{p}&\longleftrightarrow&\rho={\sum}_\kappa\,p_\kappa\,|\kappa\rangle\langle\kappa|\in\Xi\,,
\end{eqnarray}
with the obvious replacement $\mu\leftrightarrow\kappa$ given above. We note that, in spite of non-unique decomposition of states in the respective bases $\bm{p}_\mu$ and $|\kappa\rangle\langle\kappa|$, one can check by direct calculation that this mapping is well defined.~\footnote{Explicit proof will be given elsewhere. This fact also follows from the results on non-negative subtheories of quasiprobability representations considered in~\cite{WaBa12}.}

\begin{figure}[t]
\begin{center}
\includegraphics[scale=0.295]{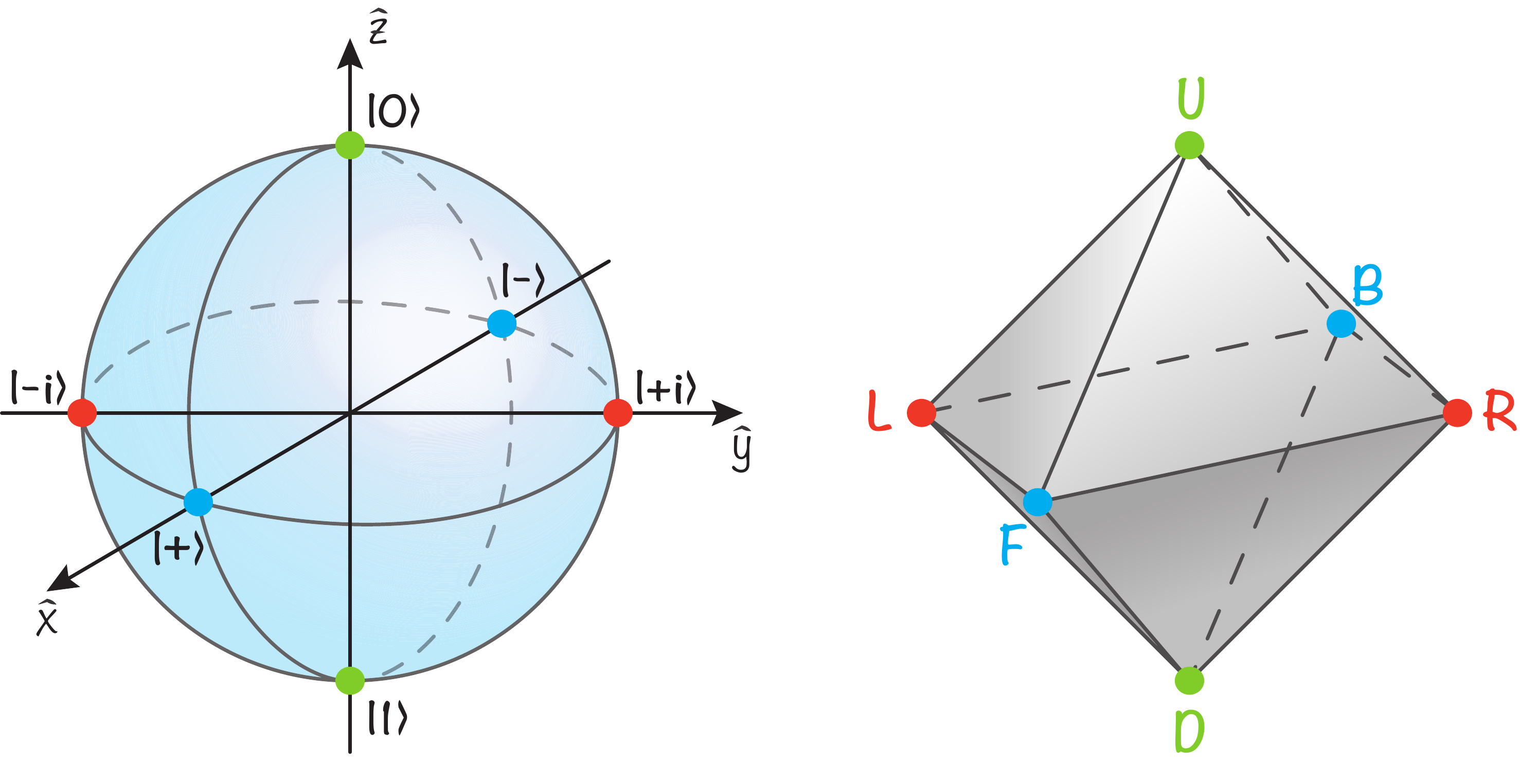}
\end{center}\vspace{-0.3cm}
\caption{\label{Fig4} On the left, the Bloch ball representing states of a qubit. On the right, the octahedron $\Xi$ providing an equivalent Hilbert space description of the toy model (corresponds bijectively to the set of epistemic states $\Delta_{\pi/2}$).}\vspace{-0.4cm}
\end{figure}

Hence, we may faithfully replace all the states $\Delta_{\pi/2}$ by their counterparts in $\Xi$. The shape of $\Xi$ is that of the octahedron inscribed in the Bloch ball representing states of a qubit, see Fig.\ref{Fig4}. It is a convex set spanned by the pure states $|\kappa\rangle\langle\kappa|$. This analogy smoothly extends to recover other aspects of the quantum formalism in our toy model. Accordingly, we get a correct account of measurements, now, described by the projectors $|\kappa\rangle\langle\kappa|$.\footnote{We note that the model of measurement can be extended in a straightforward manner to convex combinations of Pauli observables.} In this representation probabilities of outcomes are calculated via the familiar formula $P_\rho(\kappa)=\langle\kappa|\rho|\kappa\rangle$ which replaces Eq.(\ref{Probability}), and after the measurement state of the system updates according to the usual projective rule. This can be readily checked for the extremal states for which probabilities take only values $0$, $1$ or $\tfrac{1}{2}$, and by virtue of linearity directly extends to the whole set $\Xi$.
Let us further observe that a transformation $T\in \mathcal{R}_{\pi/2}$ in analogous manner rotates the cube in Fig.\ref{Fig3} as it rotates the octahedron in Fig.\ref{Fig4} (notice that these are dual solids~\cite{Ar88}, i.e. have the same set of symmetries). Hence, from the geometric picture we infer that transformations in $\mathcal{R}_{\pi/2}$ are represented by unitary maps transforming the octahedron $\Xi$ into itself, i.e. $\rho\rightarrow U\rho\, U^{\dag}$ which takes place of Eq.(\ref{Transformation}). In explicit form one recovers the Clifford group of a qubit and a  standard projective representation of $\mathcal{R}_{\pi/2}$ in $\mathbb{C}^2$, i.e. $R_{\hat{x}}(\tfrac{\pi}{2})\leftrightarrow\tfrac{1}{\sqrt{2}}\left(\mathbbm{1}-i\hat{\sigma}_{x}\right)$, $R_{\hat{y}}(\tfrac{\pi}{2})\leftrightarrow\tfrac{1}{\sqrt{2}}\left(\mathbbm{1}-i\hat{\sigma}_{y}\right)$, $R_{\hat{z}}(\tfrac{\pi}{2})\leftrightarrow\tfrac{1}{\sqrt{2}}\left(\mathbbm{1}-i\hat{\sigma}_{z}\right)$, etc. 
Thus the subset of states $\Xi$ together with the usual quantum rules provide equivalent description of the model discussed above. 

\section{Discussion}

In conclusion, we observe that the limited means of investigating the system make the toy model indistinguishable from a constrained version of a qubit. It was shown that a proper epistemic account of the system, adjusted to specific restrictions on gaining knowledge, is fully \emph{equivalent} to the description of a qubit given by the convex hull of eigenstates of Pauli operators, Clifford transformations and Pauli observables in the two-dimensional Hilbert space.

Particular interest in the model stems from a simple geometric interpretation of the underlying ontic space and the straightforward nature of epistemic restrictions imposed on the system. The latter are introduced explicitly in the description of what constitutes a measurement and how the system can be transformed. Let us emphasize a particular advantage of the \emph{two-fold} character of cognitive restrictions that we impose separately on measurements and transformations. This scheme not only gives a natural picture of how the system is processed, but also seamlessly integrates into the model state disturbance and the appropriate group theory and symmetry concepts.

Our toy model is built from the outset in the operational framework which makes clear distinction between preparation procedures, transformations and measurements. As such it befits classification of ontological models given in Ref.~\cite{HaSp10} falling into the category of $\psi$\emph{-epistemic} models. In implementing epistemic restrictions our approach is primarily focused on a conceptual mechanism and explicit construction of how restriction on gaining knowledge occurs. This should be confronted with complementary approaches seeking foundational principles like the 'knowledge balance principle' proposed in Ref.~\cite{Sp07}. We note that our model improves on the single particle aspect of the Spekkens' toy theory~\cite{Sp07} which bears only a qualitatively resemblance to a constrained qubit~\cite{CoEd11,Pu12}. It is instructive to observe that our model does not obey the 'knowledge balance principle' which boils down to the conceptual difference in implementation of epistemic restrictions in both models.
After completing this work, the author became aware that the model presented here is  equivalent to one of the models described in Ref.~\cite{WaBa12}, although the primary concern of Ref.~\cite{WaBa12} is attached to the analysis of possible non-negative quasiprobability representations, while the current model is motivated by considering restrictions on knowledge and interpretation thereof. In addition, the model presented here has a clear geometric interpretation and provides a complete operational framework of a constrained  qubit.



Being a \emph{faithful} analogue of a constrained qubit our toy model should provide a good point of departure for development of models simulating multipartite systems and further study of $\psi$-epistemic reconstructions. We note that extension of the toy model to more qubits would require incorporation of contextuality and the related concepts of entanglement and nonlocality. While acknowledging that it might constitute a difficult step we point out that explicitly operational framework of our model together with nontrivial structure of the underlying ontic state space and their transformations may be a move in the right direction.

\bibliographystyle{model1-num-names}
\bibliography{CombQuant}







\end{document}